\documentclass[aps,prl,showpacs,reprint,amsmath,amssymb,superscriptaddress] {revtex4-1}
\usepackage{graphicx}

\begin{document}

\title{Magnetic ordering at anomalously high temperatures in Dy at extreme
pressures:\ a new Kondo-lattice state?}

\author{J.~Lim}
\affiliation {Department of Physics, Washington University, St. Louis, MO 
63130, USA}

\author{G.~Fabbris}
\affiliation {Department of Physics, Washington University, St. Louis, MO 
63130, USA}
\affiliation {Advanced Photon Source, Argonne National Laboratory, Argonne, IL 
60439, USA}

\author{D.~Haskel}
\affiliation {Advanced Photon Source, Argonne National Laboratory, Argonne, IL 
60439, USA}

\author{J.~S.~Schilling}
\email[Corresponding author: ]{jss@wuphys.wustl.edu}
\affiliation {Department of Physics, Washington University, St. Louis, MO 
63130, USA}

\date{\today}

\begin{abstract}
In an attempt to destabilize the magnetic state of the heavy lanthanides Dy
and Gd, extreme pressures were applied in an electrical resistivity
measurement to 157 GPa over the temperature range 5 - 295 K. The magnetic
ordering temperature $T_{\text{o}}$ and spin-disorder resistance $R_{sd}$ of
Dy, as well as the superconducting pair-breaking effect $\Delta T_{c}$ in Y(1
at.\% Dy), are found to track each other in a highly non-monotonic fashion as
a function of pressure, all three increasing sharply above 73 GPa, the
critical pressure for a 6\% volume collapse in Dy. At 157 GPa $T_{\text{o}}$
is estimated to reach temperatures in the range 370 - 500 K, the highest
magnetic ordering temperature of any lanthanide. In contrast, $T_{\text{o}%
}(P)$ for Gd shows no such sharp increase to 105 GPa. Taken together, these
results suggest that pressures greater than 73 GPa transform Dy from a
conventional magnetic lanthanide into a Kondo lattice system with an
anomalously high magnetic ordering temperature.

\end{abstract}

\maketitle

Subjecting a solid to arbitrarily high pressures will successively break up
its atomic shell structure, leading to a rise and fall in all condensed matter
properties, including magnetism and superconductivity, until finally only a
structureless Thomas-Fermi gas remains \cite{jim1}. Although such astronomic
pressures are not available in the laboratory, recent technological
developments do allow measurements of the magnetic and superconducting
properties of matter to multi-megabar pressures where the increase in energy
(1-10 eV/atom) is sufficient to significantly alter electronic states. Systems
with magnetic instabilities exhibit some of the most fascinating properties in
current condensed matter physics, including topological insulators
\cite{beidenkopf}, Kondo lattice behavior \cite{aynajian}, and exotic forms of
superconductivity \cite{scalapino}. With the availability of extreme
pressures, it may now be possible to transport many conventional magnetic
systems to ones exhibiting new and unexpected magnetic and/or superconducting properties.

Due to the high degree of localization of their 4$f$ orbitals, the heavy
lanthanide metals, such as Dy, display the purest form of local moment
magnetism. It can be estimated that the molar volume of the heavy lanthanides
would have to be compressed approximately five-fold before the
nearest-neighbor overlap of 4$f$ orbitals becomes sufficient to prompt a
local-to-itinerant transition \cite{jim}. Other forms of magnetic instability
may require less compression. Jackson \textit{et al.} \cite{jackson1} have
pointed out that the heavy lanthanides Gd, Tb, Dy, Ho, Er, and Tm exhibit
conventional magnetic ordering to pressures of $\sim10$ GPa as evidenced by
the fact that their respective magnetic ordering temperatures $T_{\text{o}}$
obey de Gennes factor scaling \cite{degennes}. Were the 4$f$ magnetic state to
become unstable under extreme pressure, such scaling would \textit{not} continue.

Dy is a trivalent heavy lanthanide with hcp structure, a 4$f^{\text{9}}$
electron configuration, antiferromagnetism below 178 K, and ferromagnetism
below 85 K \cite{behrendt1}. In this paper we present the results of
temperature-dependent dc electrical resistivity measurements on Dy to
pressures as high as 157 GPa, well above the pressure of 73 GPa where Dy
suffers a 6\% volume collapse, reportedly transforming from the hexagonal hR24
to a body-centered monoclinic (bcm) structure \cite{patterson1}. As the
applied pressure passes through 73 GPa, the magnetic ordering temperature
$T_{\text{o}}$ begins to increase dramatically, appearing to rise well above
ambient temperature to values surpassing those for any lanthanide. These and
parallel resistivity studies on both Gd metal and the dilute magnetic alloys
Y(1 at.\% Dy) and Y(0.5 at.\% Gd) give evidence that extreme pressures
transform Dy into a Kondo lattice system with a significantly enhanced
magnetic ordering temperature $T_{\text{o }}$lying between 370~K and 500 K at
157 GPa.

Resistivity samples were cut from Dy and Gd foil (99.9\% Alfa Aesar). The
dilute magnetic alloys were prepared by argon arc-melting stoichiometric
amounts of Y (99.9\% Ames Lab \cite{ames}) with Dy or Gd dopant. Following the
initial melt, the sample was turned over and remelted several times with less
than 0.1\% weight loss.

\begin{figure}[t]
\includegraphics[width = 8.5 cm]{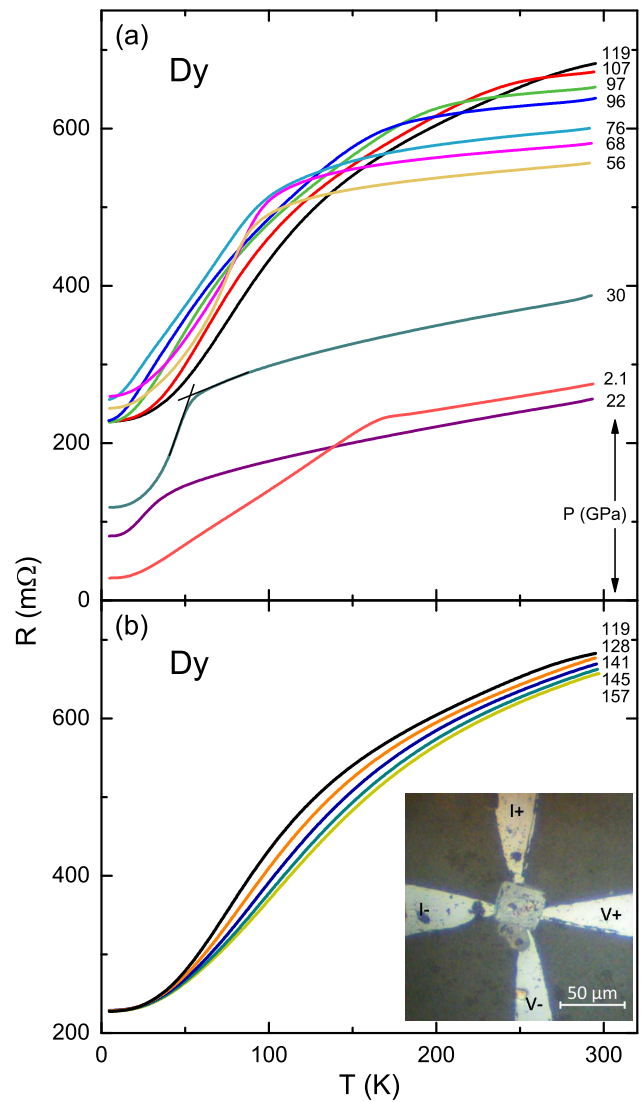} \caption{\label{fig1}(color 
online) Resistance of Dy versus temperature to 295 K in pressure
ranges: \ (a) 2.1 - 119 GPa, and (b) 119 - 157 GPa. The magnetic ordering
temperature $T_{\text{o}}$ is determined by the intersection point of two
straight lines, as illustrated in Fig. \ref{fig1}(a) at 30 GPa. Inset to (b) shows
four-point dc resistivity technique used. Dy sample $(30\times30\times5$ $\mu
$m$^{3})$ rests on four Pt leads (4 $\mu$m thick) on insulated Re
gasket.}
\end{figure}

To generate pressures well beyond the volume collapse pressure of Dy at 73
GPa, a diamond anvil cell (DAC) made of CuBe alloy was used \cite{Schilling84}%
. Three separate non-hydrostatic high-pressure experiments on Dy were carried
out. In run 1 two oppposing diamond anvils (1/6-carat, type Ia) with 0.5 mm
diameter culets were used. In runs 2 and 3 the anvils had a 0.35 mm diameter
culet beveled at 7$^{\circ}$ to a 0.18 mm central flat. The Re gasket (6-7 mm
diameter, 250 $\mu$m thick) was preindented to 30 $\mu$m and a 80 $\mu$m
diameter hole electro-spark drilled through the center (for the 0.5 mm culet
anvils the gasket was preindented to 80 $\mu$m with a 250 $\mu$m diameter
hole). The center section of the preindented gasket surface was filled with a
4:1 cBN-epoxy mixture to insulate the gasket and serve as pressure medium (see
inset in Fig. \ref{fig1}). The thin Dy sample (dimensions $\sim$30$\times$30$\times$5
$\mu$m$^{3}$) was then placed on top of four thin Pt leads for a four-point dc
electrical resistivity measurement. Two experimental runs were carried out on
the thin Gd sample (same dimensions as for Dy) using beveled anvils as above.
Further details of the non-hydrostatic high pressure resistivity technique are
given in a paper by Shimizu \textit{et al}. \cite{Shimizu05}.

A He-gas-driven membrane was utilized to change pressure at any temperature
above 3 K \cite{daniels1}. In the measurement on the Y(1 at.\% Dy) alloy, one
ruby sphere was positioned at the center of, and another directly next to, the
sample. The average pressure over the sample was determined \textit{in situ}
at 25 K with the standard ruby fluorescence technique using the revised
pressure scale of Chijioke \textit{et al}. \cite{Chijioke05}. In the
resistivity measurements on Dy and Gd, pressure was determined using both ruby
fluorescence and, in the upper pressure range, Raman spectroscopy from the
frequency shift of the diamond vibron \cite{raman1}. The \textquotedblleft
home-made\textquotedblright\ Raman spectrometer utilizes a Nikon
metallographic microscope coupled fiber-optically to a sensitive QE65000
spectrometer from Ocean Optics \cite{benzene}. The values of the pressures
given are averaged over the sample to an estimated accuracy of $\pm10\%.$ In
these experiments temperatures as low as 1.3 K were reached in an Oxford flow
cryostat. Further experimental details of the DAC and cryostat are given
elsewhere \cite{Schilling84,klotz1,debessai1}.

The present resistivity studies on Dy were carried out in three separate
experiments. In Fig. \ref{fig1} the electrical resistance $R(T)$ of Dy from run 3 is
plotted versus temperature at 14 different pressures to 157 GPa. The magnetic
ordering temperature $T_{\text{o}}$ is defined by the kink in the $R(T)$
dependence clearly seen near 170 K at 2.1 GPa, the lowest pressure of the
experiment. At higher pressures this kink broadens somewhat due to pressure
gradients across the sample, but remains clearly visible to 107 GPa. We define
$T_{\text{o}}$ by the intersection point of two straight lines, as illustrated
for the data at 30 GPa in Fig. \ref{fig1}(a).

Fig. \ref{fig1}(b) shows the resistance data $R(T)$ on Dy at the highest pressures. The
fact that the $R(T)$ curves continue to shift to higher temperatures with
pressure indicates that the magnetic ordering temperature $T_{\text{o}}$ has
increased above 295 K (see Supplemental Material \cite{suppl} for a detailed
analysis). The residual resistance $R(5$ K) initially increases appreciably
with pressure as defects are introduced into the sample through plastic
deformation by the non-hydrostatic pressure. However, for pressures of 56 GPa
and above the pressure cell stabilizes, the relatively small changes (both
positive and negative) in $R(5$ K) at higher pressures likely arising from
small displacements of the voltage contacts.

\begin{figure}[t]
\includegraphics[width = 7.9 cm]{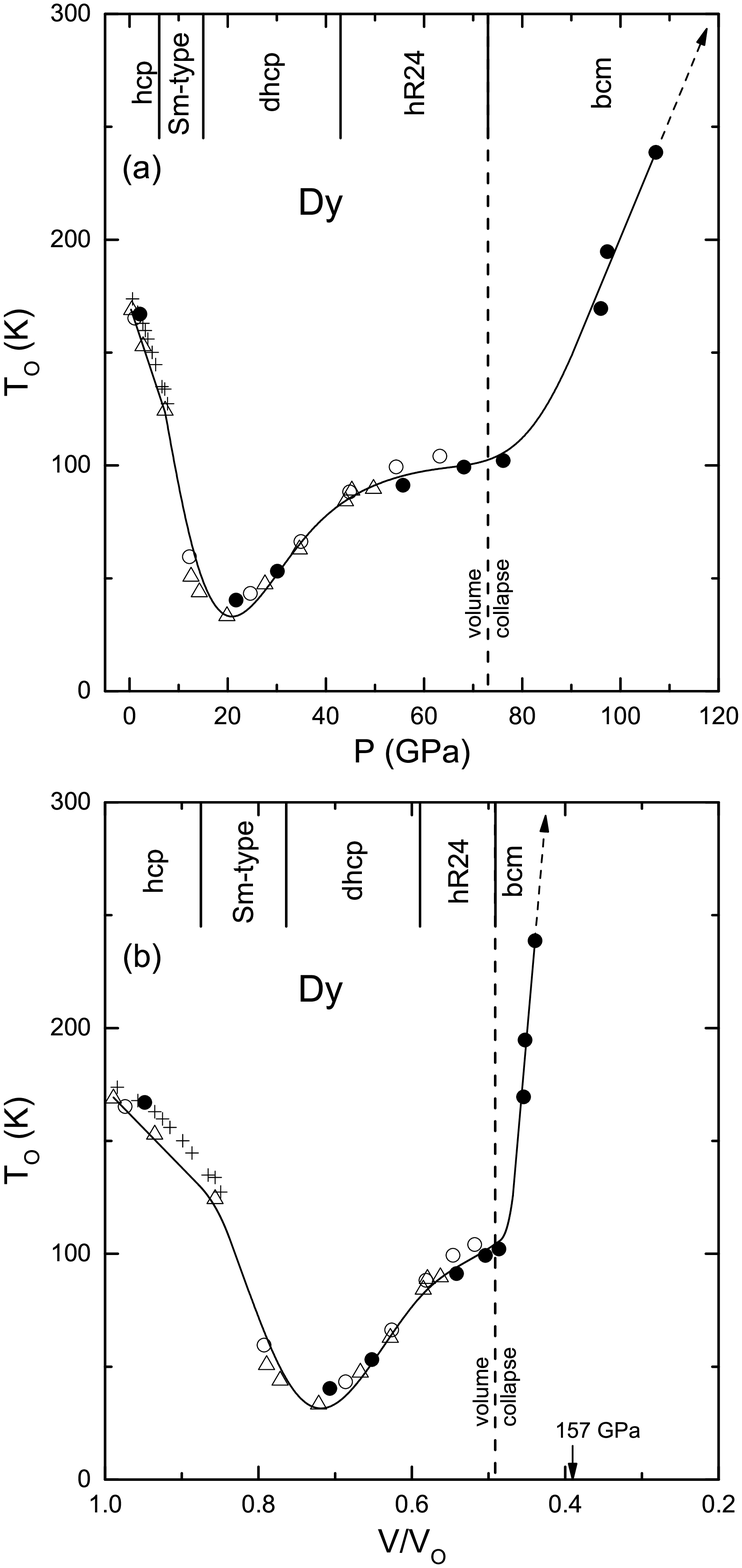} \caption{\label{fig2} Magnetic ordering temperature $T_{\text{o}}$ of Dy versus
(a) pressure or (b) relative volume $V/V_{o}$ using equation of state from Ref
\cite{patterson1}. ($+$) earlier studies to 7.4 GPa with slope $dT_{\text{c}%
}/dP=-6.7$ K/GPa \cite{jackson1}; ($\triangle$) present resistivity
measurements in run 1 with initial slope $-6.5$ K/GPa , ({\large o}) run 2,
($\bullet$) run 3. Vertical dashed line marks pressure of volume collapse for
Dy at 73 GPa. Crystal structures at top of graph are for Dy \cite{patterson1}.
In both plots the extended solid line through data points is guide to the
eye.}
\end{figure}

In Fig. \ref{fig2}(a) $T_{\text{o}}$ is plotted versus pressure to 107 GPa for all
three experiments on Dy. The results are in reasonable agreement both with
earlier results of Jackson \textit{et al.} \cite{jackson1} to 7.4 GPa and very
recent resistivity studies of Samudrala \textit{et al.} \cite{samu} to 69 GPa.
The pressure dependence $T_{\text{o}}(P)$ is seen to be highly non-monotonic,
presumably in response to changes in crystal structure given at the top of the
graph. $T_{\text{o}}(P)$ initially decreases with pressure, the rate of
decrease roughly doubling at the hcp to Sm-type phase transition before
passing through a minimum and increasing at $\sim20$ GPa near the transition
to the dhcp structure. The rate of increase is then diminished near the
transition to hexagonal hR24 \cite{patterson1,shen}.

Particularly intriguing is the dramatic increase in slope $dT_{\text{o}}/dP$
following the hR24 to body-centered monoclinic (bcm) transition at 73 GPa
\cite{patterson1}. Above 73 GPa, $T_{\text{o}}(P)$ for Dy begins to increase
rapidly with pressure and appears to pass through 295 K near 119 GPa.
Extrapolating the straight line in Fig. \ref{fig2}(a) above 100 GPa to the highest
pressure of 157 GPa yields the estimate $T_{\text{o}}\approx$ 500 K. How
dramatic this increase in $T_{\text{o}}$ really is can be seen in Fig. \ref{fig2}(b)
where $T_{\text{o}}$ is plotted versus relative sample volume $V/V_{\text{o}}
$, a parameter with a more direct physical significance than pressure $P$. In
Fig. \ref{fig2}(b) the rate of increase of $T_{\text{o}}$ below $V/V_{\text{o}}%
\simeq0.51$ (above 73 GPa), is seen to be \textit{much} steeper than the
initial rate of decrease of $T_{\text{o}}$\ near 0 GPa where $V/V_{\text{o}%
}=1$. Extrapolating $T_{\text{o }}$in Fig. \ref{fig2}(b) linearly to $V/V_{\text{o}%
}=0.39$ (157 GPa), yields the estimate $T_{\text{o}}\approx$ 430 K. In the
Supplemental Material \cite{suppl} the resistance data $R(T)$ for $P>96$ GPa
are analyzed in detail, yielding the semi-quantitative value $T_{\text{o}%
}\approx$ 370 K. At 157 GPa the magnetic ordering temperature of Dy is thus
estimated to lie in the range 370 - 500 K. This is the highest magnetic
ordering temperature of any lanthanide, surpassing Gd's ambient pressure value
$T_{\text{o}}\simeq$ 292 K \cite{colvin1}.

 \begin{figure}[t]
\includegraphics[width = 8.5 cm]{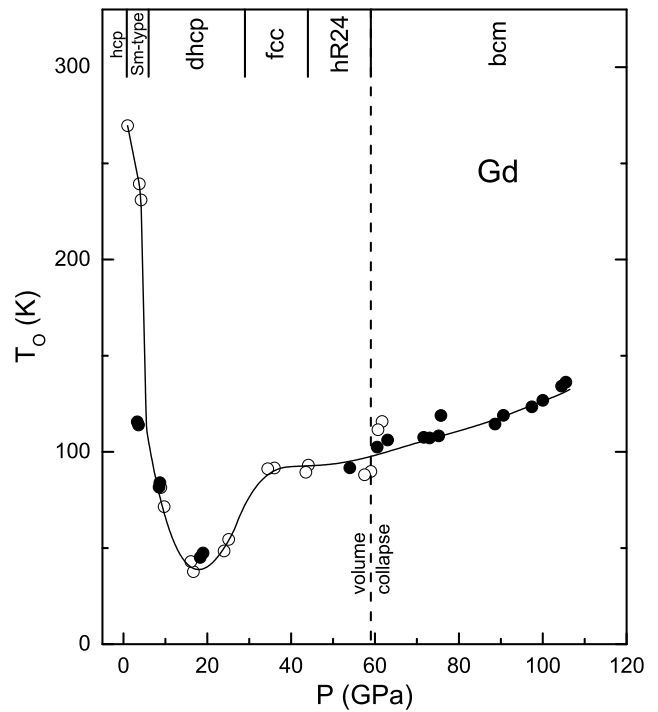} \caption{\label{fig3} Magnetic ordering temperature $T_{\text{o}}$ of Gd
versus pressure to 105 GPa. Vertical dashed line marks pressure of volume
collapse for Gd at 59 GPa. Crystal structures at top of graph are from Ref
\onlinecite{gd}.}
\end{figure}

An obvious question remains: \ what is(are) the mechanism(s) responsible for
the highly non-monotonic dependence of the magnetic ordering temperature
$T_{\text{o}}\ $of Dy on pressure, particularly its dramatic increase just
above 73 GPa? To shed some light on this matter, it is interesting to compare
the pressure dependence $T_{\text{o}}(P)$ for Dy in Fig. \ref{fig2}(a) to that for its
lighter next-nearest-neighbor lanthanide, Gd, shown in Fig. \ref{fig3}. A detailed
comparison of the $T_{\text{o}}(P)$ data for Dy and Gd reveals striking
similarities to 70 GPa. Fleming and Liu \cite{fleming} show that the initial
decrease in $T_{\text{o}}\ $with pressure for Gd, Tb, and Dy is a direct
consequence of shifts in the electronic energy bands. Gd \cite{gd} and Dy
\cite{patterson1} undergo a very similar set of structural phase transitions
(those for Gd at somewhat reduced pressures) driven by increasing 5\textit{d}
electron occupation with pressure \cite{pettifor}. The highly non-monotonic
dependence $T_{\text{o}}(P)$ seen for both Gd and Dy to 70 GPa is thus likely
driven by shifts in the energy bands and changes in the Fermi surface
associated with the multiple structural phase transitions.

 \begin{figure}[t]
\includegraphics[width = 8.5 cm]{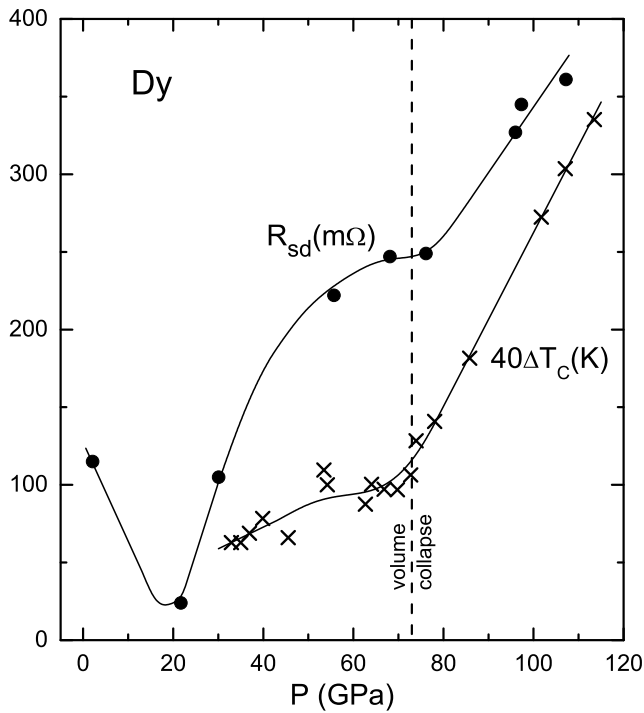} \caption{\label{fig4} Pressure dependence of both the estimated spin-disorder
resistance $R_{sd}$ at 295 K and $\Delta T_{c}$, the reduction in the value of
the superconducting transition temperature of Y(1 at.\% Dy) compared to that
of pure Y. Note that the value for $\Delta T_{c}$ has been enhanced
40$\times.$}
\end{figure}

For pressures above 70 GPa, however, the pressure dependences of $T_{\text{o}%
}(P)$ for Gd and Dy are seen to differ significantly. Whereas $T_{\text{o}%
}(P)$ for Dy displays a sharp upturn above 73 GPa, that for Gd increases only
gradually over the entire pressure range 40 - 105 GPa with no sign of a rapid
upturn near the volume collapse pressure for either element. Since the phase
diagrams of Gd and Dy are so similar, the sharp upturn in $T_{\text{o}}(P)$
for Dy almost certainly does not originate from shifts in the energy bands but
rather from an anomalous increase in the effective exchange interaction $J$
between the magnetic ions and the conduction electrons. Unlike the
conventional lanthanide Gd, Dy appears to enter a new state for pressures
above 73 GPa, the nature of which we now explore.

A long-standing strategy \cite{matthias,maple2} to probe the magnetic state of
a given ion is to alloy the magnetic ion in dilute concentration with a
superconductor and determine to what extent the superconducting transition
temperature is suppressed $\Delta T_{\text{c}}.$ The pressure dependence
$T_{\text{c}}(P)$ for Y(0.5 at.\% Gd) was found to faithfully track that for Y
to the maximum pressure of 126 GPa \cite{fabbris1}. This gives evidence that
over this pressure range Gd remains a conventional magnetically stable
lanthanide. The absence of magnetic instabilities in Gd, even at extreme
pressures, is not surprising since the magnetic state of Gd with its
half-filled 4$f^{7}$ shell is the most stable of all elements, its 4$f^{7}$
level lying $\sim9$ eV below the Fermi level \cite{yin}.

This contrasts with the $T_{\text{c}}(P)$ data for Y and its dilute magnetic
alloy Y(1 at.\% Dy) where the obtained pressure dependence of $\Delta
T_{\text{c}}$ is plotted in Fig. \ref{fig4} (see Supplemental Material for details
\cite{suppl}). As for $T_{\text{o}}$, the pressure dependence of $\Delta
T_{c}$ is seen to show a sharp upturn above 73 GPa, reaching the value $\Delta
T_{\text{c}}\approx$ 9 K at the highest pressure. Such a dramatic suppression
of superconductivity strongly suggests Kondo pair breaking, implying that the
normally \textit{positive} exchange interaction $J$ becomes \textit{negative}
for $P>$ 73 GPa, a pre-requisite for Kondo effect phenomena. A negative value
of $J$ for $P>73$ GPa in Dy signals that Dy has been transformed into a Kondo
lattice system.

What is the mechanism behind the sharp upturn in $T_{\text{o}}$ and $\Delta
T_{c}$ above 73 GPa? When $J$ is negative, the covalent mixing interaction
between the 4$f$ and conductions electrons is dominant. $J$ then depends on
the mixing matrix element $V_{sf}$ and the 4$f$-electron stabilization energy
$E_{ex}$ according to $J\propto-\left\vert V_{sf}\right\vert ^{2}/E_{ex},$
where $E_{ex}$ is assumed small compared to the Coulomb repulsion $U$ between
electrons on the same orbital \cite{schrieffer}. As the magnetic ion heads
toward the mixed-valence state with increasing pressure, $E_{ex}$ approaches
zero and/or $V_{sf}$ increases. In either case $\left\vert J\right\vert $ is
enhanced. The magnetic ordering temperature $T_{\text{o}}\propto J^{2}$ would
thus both be expected to increase with pressure until $\left\vert J\right\vert
$ becomes so large that the local magnetic moment begins to be compensated
through the exponentially increasing Kondo spin screening, as anticipated in
the simple Kondo-lattice model \cite{doniach1,yang}. This could then lead to
an anomalously high value of $T_{\text{o}}$, such as observed for Dy at
extreme pressure, a value perhaps surpassing that possible for normal positive
exchange interactions.

Interestingly, in Fig. \ref{fig4} it is also seen that the spin-disorder resistance
$R_{sd}(P)$ at 295 K (see Supplemental Material \cite{suppl}) also changes
under pressures to 105 GPa in a manner similar to that seen for $T_{\text{o}%
}(P)$ in Fig. \ref{fig2}(a). That in Dy all three quantities $T_{\text{o}}$, $\Delta
T_{\text{c}}$, and $R_{sd},$ track each other as a function of pressure is not
surprising since, in the simplest model, all three are proportional to the
square of the exchange interaction $J^{2}.$

In summary, measurements of the electrical resistivity of Dy metal to extreme
pressures reveal that the magnetic ordering temperature $T_{\text{o}} $
exhibits a highly non-monotonic pressure dependence, rising dramatically for
$P>$ 73 GPa to unprecedentedly high values in the range 370 - 500 K. Parallel
experiments on Gd and dilute magnetic alloys of Gd and Dy with Y give evidence
that under extreme pressures Dy is transformed from a magnetically
conventional lanthanide into a dense Kondo lattice system with anomalously
high values of $T_{\text{o}}$. At pressures even higher than those in the
present experiment, $T_{\text{o}}(P)$ would be expected to pass through a
maximum and fall rapidly to 0 K at a quantum critical point
\cite{doniach1,yang}. A search for further lanthanide and actinide systems
with anomalously high magnetic ordering temperatures would be of considerable
interest and is underway.

\begin{acknowledgments}
The authors would like to thank T. Matsuoka
and K. Shimizu for sharing information on their high-pressure electrical
resistivity techniques used in this study. Thanks are due A. Gangopadhyay for
his critical reading of the manuscript. This work was supported by the
National Science Foundation (NSF) through Grant No. DMR-1104742 and by the
Carnegie/DOE Alliance Center (CDAC) through NNSA/DOE Grant No.
DE-FC52-08NA28554. Work at Argonne National Laboratory is supported by the
U.S. Department of Energy, Office of Science, under contract No. DE-AC02-06CH11357.
\end{acknowledgments}

\end{document}